\documentclass[prl,twocolumn,superscriptaddress,nofootinbib]{revtex4-1}
\usepackage{amsmath,amssymb,graphicx,amsfonts}
\usepackage{color, cancel}
\usepackage{subfigure}
\pdfoutput=1

\begin{document}
\newcommand{\fp}[2]{\frac{\partial #1}{\partial #2}}
\newcommand{\dfp}[2]{\dfrac{\partial #1}{\partial #2}}
\newcommand{\fpp}[3]{\frac{\partial^2 #1}{\partial #2 \partial #3}}
\newcommand{\norm}[1]{\left\lVert #1 \right\rVert}

\newcommand*\diff{\mathop{}\!\mathrm{d}}
\newcommand*\Diff[1]{\mathop{}\!\mathrm{d^#1}}
\newcommand{\be}{\begin{equation}}
\newcommand{\ee}{\end{equation}}
\newcommand{\ba}{\begin{eqnarray}}
\newcommand{\ea}{\end{eqnarray}}

\title{Bayesian naturalness of the CMSSM and CNMSSM}



\author{Doyoun Kim}
\affiliation{ARC Centre of Excellence for Particle Physics at the Tera-scale, School of Physics, Monash University, Clayton, VIC 3800, Australia}

\author{Peter Athron}
\affiliation{ARC Centre of Excellence for Particle Physics at the Terascale,
School of Chemistry and Physics, University of Adelaide, Adelaide, Australia} 

\author{Csaba Bal\'azs}
\author{Benjamin Farmer}
\author{Elliot Hutchison}
\affiliation{ARC Centre of Excellence for Particle Physics at the Tera-scale, School of Physics, Monash University, Clayton, VIC 3800, Australia}


\date{\today}

\begin{abstract}

The recent discovery of the $125.5$ GeV Higgs boson at the LHC has fuelled interest in the Next-to-Minimal Supersymmetric Standard Model (NMSSM) as it may require less fine-tuning than the minimal model to accommodate such a heavy Higgs.  To this end we present Bayesian naturalness priors to quantify fine-tuning in the (N)MSSM.  These priors arise automatically as Occam razors in Bayesian model comparison and generalize the conventional Barbieri-Giudice measure.  In this paper we show that the naturalness priors capture features of both the Barbieri-Giudice fine-tuning measure and a simple ratio measure that has been used in the literature.  We also show that according to the naturalness prior the constrained version of the NMSSM is less tuned than the CMSSM.


\end{abstract}

\pacs{}

\maketitle

\section{Introduction}

Naturalness is a guiding principle in search of new physics beyond the Standard Model (SM) \cite{Giudice:2013yca}.  A naturalness problem arises in the SM since the Higgs mass is sensitive to new physics above the electroweak scale and only delicate fine-tuning amongst the fundamental parameters can stabilize it.
Supersymmetry cancels the quadratic divergence in the Higgs mass improving naturalness.  In the Minimal Supersymmetric Standard Model (MSSM), however, the large radiative corrections that lift the Higgs mass reintroduce some fine-tuning \cite{Giudice:2013yca}.

The recent Higgs discovery makes the little hierarchy problem more acute \cite{Aad:2012tfa, Chatrchyan:2012ufa}.  This triggered interest in supersymmetric models that can naturally accommodate a 125.5 GeV Higgs, such as the Next-to-Minimal Supersymmetric Standard Model (NMSSM).  A new $F$-term in the NMSSM, proportional to the Higgs-singlet coupling $\lambda$, boosts the tree level Higgs mass. 
Natural NMSSM scenarios have been presented where $\lambda$ remains perturbative up to the GUT scale \cite{King:2012tr}, and in $\lambda$-SUSY scenarios where $\lambda$ is only required to remain perturbative up to a scale just above TeV \cite{Gherghetta:2012gb}.

To show that the NMSSM is less fine-tuned than the MSSM one has to quantify naturalness.  Conventional fine-tuning measures rely on the sensitivity of the weak scale to changes in the fundamental parameters of the model.  In Bayesian model comparison such measure arises automatically as a Jacobian of the variable transformation from the Higgs vacuum expectation values (VEVs) to the fundamental parameters \cite{Allanach:2007qk, Cabrera:2008tj, Cabrera:2009dm, Ghilencea:2012gz, Ghilencea:2012qk, Fichet:2012sn}.  

In this paper we present the NMSSM fine-tuning prior.  We examine how the prior varies with the parameter of the constrained NMSSM and compare it to the Barbieri-Giudice measure \cite{Ellis:1986yg, Barbieri:1987fn} and the simple ratio measure \cite{Baer:2012up, Baer:2012cf, Baer:2013bba, Baer:2013gva}.       
In a longer companion paper we will provide the full details of the derivation of the presented Jacobian and carry out a detailed numerical analysis for the unconstrained NMSSM.

\section{Measuring fine-tuning}
Naturalness of supersymmetric models is quantified in various different ways in the literature today.  One of the simplest fine-tuning measures is \cite{Baer:2012up, Baer:2012cf, Baer:2013bba, Baer:2013gva}
\begin{equation}
	\Delta_{EW}=\max \{|C_i|/(m_Z^2/2)\} \label{del_EW} ,
\end{equation}
which is based on the electroweak symmetry breaking (EWSB) condition of the MSSM
\begin{equation}
	\frac{m_Z^2}{2}=\frac{(m_{H_d}^2+\delta m_{H_d}^2)-(m_{H_u}^2+\delta m_{H_u}^2)\tan^2\beta}{\tan^2\beta-1}-\mu^2 \label{Eq:mzsq},
\end{equation}
where $\delta m_{H_u}^2$ and $\delta m_{H_d}^2$ are the one loop tadpole corrections to the tree level minimisation conditions.
The $C_i$ ($i$ = $m_{H_u}^2$, $m_{H_d}^2$, $\delta m_{H_u}^2$, $\delta m_{H_d}^2$, $\mu^2$) in Eq.~(\ref{del_EW}) are the additive terms appearing in Eq.~(\ref{Eq:mzsq}), specified by the index. 

An alternative and widely used measure of naturalness, the Barbieri-Giudice measure \cite{Ellis:1986yg, Barbieri:1987fn}, accounts for correlations between the terms in Eq.~(\ref{Eq:mzsq}), 
\begin{equation}
	\Delta_{BG}(p_i)=\left|\fp{\ln m_Z^2}{\ln p_i^2}\right|, \;\;\;  \Delta_{BG} = \max \{ \Delta_{BG}(p_i) \}.
\label{del_BG}
\end{equation}
where $p_i$ are the input parameters of the model, for some chosen parametrisation. Alternatively some authors combine the $\Delta_{BG}(p_i)$ by summation in quadrature: $\tilde{\Delta}_{BG}^2=\sum_i{\Delta^2_{BG}(p_i)}$.
The Barbieri-Giudice measure quantifies the sensitivity of the observable $m_Z^2$ to the parameters $\{p_i\}$, e.g. $\Delta_{BG}(p_i)=10$, means a 1 percent change in $p_i$ leads to 10 percent change in $m_Z^2$.  

Alternative measures have also been proposed \cite{Anderson:1994tr, Anderson:1994dz, Athron:2007ry} but these are not considered here.



In Bayesian model comparison the Barbieri-Giudice measure arises automatically as the special case of a more general fine-tuning measure \cite{Allanach:2007qk, Cabrera:2008tj, Cabrera:2009dm, Ghilencea:2012gz, Ghilencea:2012qk, Fichet:2012sn}.  In this framework the odds ratio between competing models is defined in terms of the ratio of marginal likelihoods, or evidences
\begin{equation}
	\mathcal{E}(\mathcal{D},\mathcal{M}) = \int_{\Omega} \mathcal{L}_\mathcal{D}(p_i;\mathcal{M}) \pi(p_i|\mathcal{M}) \,d^{N}\!p_i .
\label{Eq:evidence}
\end{equation}
Here $\mathcal{L}_\mathcal{D}$ is the likelihood function for the data $\mathcal{D}$, quantifying the goodness of fit of the model $\mathcal{M}$ to the data at each point in the model's $N$ dimensional parameter space $\{p_i\}$.  The distribution $\pi$ assigns a probability density to each parameter space point as assessed prior to the data $\mathcal{D}$ being learned, and $\Omega$ is the domain over which this pdf is non-zero. 

However, for computing likelihoods, and for scanning, the parameter set $\{p_i\}$ on which $\pi$ is most sensibly defined is often less convenient to work with than a set of derived parameters or ``observables'' $\mathcal{O}_i$, some of which, such as $m_Z$, may be precisely measured. Switching to these new variables distorts the prior density, as quantified by the Jacobian of the transformation,
\begin{equation}
	d^N\!p_i=\left|\fp{p_i}{\mathcal{O}_j}\right|d^N\mathcal{O}_j .
        \label{Eq:ptoO}
\end{equation}
In the new coordinates the sharply known observables can be easily marginalised out, reducing the dimension of the parameter space. Choosing logarithmic priors on $\{p_i\}$ and neglecting constants which divide out of evidence ratios gives
\begin{equation}
	\mathcal{E}(\mathcal{D},\mathcal{M}) = \frac{1}{V_{\log}}\int_{\Omega'} \mathcal{L}_{\mathcal{D}'} \left.\frac{1}{\Delta_J}\right|_{\hat{\mathcal{O}}=\hat{\mathcal{D}}} \frac{d^{N-r}\mathcal{O}'_j}{\mathcal{O}'_j}
\label{bayestest}
\end{equation}
where $\hat{\mathcal{O}}$ and $\hat{\mathcal{D}}$ are the $r$ observables and data used in the dimensional reduction, $\mathcal{D'}$ and $\mathcal{O'}$ are the data and observables remaining, $V_{\log} = \int_\Omega d^N\!p_i/p_i$ is the ``logarithmic'' volume of the parameter space in the original coordinates, $\Omega'$ is the part of $\Omega$ orthogonal to the removed dimensions and
\begin{equation}
	\Delta_{J}=\left|\fp{\ln\mathcal{O}_j}{\ln p_i}\right|.
\end{equation}
If log priors are not used extra terms will also appear. $\Delta_J^{-1}$ appears in the evidence through the transformation of the prior density to new coordinates, $\pi(\mathcal{O}_j)=\Delta_J^{-1}(p_i/\mathcal{O}_j)\pi(p_i)=\Delta_J^{-1}(1/\mathcal{O}_j)$.  The last equality follows from initially choosing log priors (neglecting normalisation constants)\footnote{For brevity the single parameter form is written here, but the generalisation is straightforward.}. One can then scan the derived parameters using log priors with $\Delta_J^{-1}$ as an ``effective'' prior weighting of the likelihood, and obtain a posterior weighting of points compatible with the original prior. 

Clearly {\em all} parameters do not have to be exchanged for observables. When a single parameter $p^2$ is exchanged for $\mathcal{O}_i = m_Z^2$, $\Delta_J$ is the Barbieri-Giudice sensitivity, $\Delta_{BG}(p_i)$, in Eq.~(\ref{del_BG}). In general more than one low energy observable is involved in the transformation, so the relevant Jacobian contains more structure than the Barbieri-Giudice measure. 

For example most MSSM spectrum generators take $(m_Z,\tan\beta,m_t)$ as input instead of $(\mu,B,y_t)$; the transformation $(\mu,B,y_t)\to (m_Z,\tan\beta,m_t)$ thus emerges as sensible choice which can quantify unnatural cancellations required to keep $m_Z \ll M_{SUSY}$. The resulting Jacobian
\begin{equation}
\begin{aligned}
	\Delta_J^\textrm{CMSSM}=&\left|\fp{\ln(m_Z^2,\tan\beta,m_t^2)}{\ln(\mu_0^2,B_0,y_0^2)}\right| \\
           =&\left(\frac{M_Z^2}{2\mu^2}\frac{B}{B_0}\frac{\tan^2\beta-1}{\tan^2\beta+1}\fp{\ln y_t^2}{\ln y_0^2} \right)^{-1}          
        \label{del_JCMSSM},
\end{aligned}
\end{equation}
automatically includes $\Delta_{BG}(\mu_0,B_0,y_0)$ as a single column (where the subscript $0$ denotes the GUT scale parameter value). The extra columns of $\Delta_J$ account correctly for correlations between the $m_Z$ related tunings and those coming from the Higgs VEVs and top mass. The Yukawa RGE factor $\fp{\ln y_t^2}{\ln y_0^2}$ is constant over the CMSSM parameter space at the 1-loop level and so we neglect it. It is close to one anyway so the constant shift this induces in the logarithms of tunings reported in our numerical analysis is very small.

Importantly, we see that Eq.~(\ref{bayestest}) captures much of the intuition behind the fine-tuning problem. We see that to be preferred in a Bayesian test, a model needs to have overlapping regions of both high likelihood and low fine-tuning (and that this region should not be too small relative to the prior volume $V_{\log}$, which is itself a naturalness-style requirement). 

The extension of $\Delta_{J}$ to the NMSSM goes as follows.  As indicated above, $\Delta_J$ in practice depends on the particular spectrum generator of choice as well as the definition of the model.  For concreteness we consider a constrained version of the NMSSM (CNMSSM)\footnote{In the literature the definition of the CNMSSM varies.  Sometimes CNMSSM refers to the model with full scalar universality and when this constraint is relaxed like in our case it is called the semi-constrained NMSSM.}, defined at the GUT scale to have a universal gaugino mass, $M_{1/2}$; a universal soft trilinear mass, $A_0$ and all MSSM-like soft scalar masses equal to $m_0$, but the new soft singlet mass, $m_S$ is left unconstrained at the GUT scale. Thus the model has the parameter set $(M_0, M_{1/2}, A_0, \lambda_0, \kappa_0, m_S)$ which can be compared to the CMSSM set of  $(M_0, M_{1/2}, A_0, \mu_0, B_0)$.   

The effective prior weighting we present is chosen to be suitable for Bayesian studies with numerical implementation in both the spectrum generator \texttt{NMSPEC} in the \texttt{NMSSMTools 4.1.2} package and the newly developed spectrum generator \texttt{Next-to-Minimal SOFTSUSY} \cite{Allanach:NMSOFTSUSY} distributed with \texttt{SOFTSUSY 3.4.0}.  For a constrained model as defined above the spectrum generators trade  $(\lambda_0,\kappa_0,m_S^2)$ for $(\lambda,m_Z,\tan\beta)$ giving the user the input parameters of $(M_0, M_{1/2}, m_Z, \tan\beta, A_0, \lambda)$, i.e. just $\lambda$ in addition to the usual CMSSM inputs used in spectrum generators.  

 
This transformation gives rise to a Jacobian,
\begin{equation}
 d\lambda_0 d\kappa_0 dm_{S_0}^2 = J_{\mathcal{T}_0} d\lambda dM_z^2 d\tan\beta 
\end{equation}
which may be written as,
\begin{equation}
\begin{aligned}       
	J_{\mathcal{T}_0} &= J_{\mathcal{T}_{\kappa m_S}^\lambda} J_{RG} \\
        &=
          \begin{vmatrix}
		&\fp{\kappa}{m_Z^2}   &\fp{m_S^2}{m_Z^2}&\\
		&\fp{\kappa}{\tan\beta}&\fp{m_S^2}{\tan\beta}&
	  \end{vmatrix}_\lambda
          \begin{vmatrix}
		&\fp{\lambda_0}{\lambda}&\fp{\kappa_0}{\lambda}&\\
		&\fp{\lambda_0}{\kappa} &\fp{\kappa_0}{\kappa}&
          \end{vmatrix}
          \left|\fp{m_{S_0}^2}{m_{S}^2}\right|
\label{Eq:Jac_kapms}
\end{aligned}
\end{equation}
The Jacobian $J_{\mathcal{T}_{\kappa m_S}^\lambda}$ can be rewritten in terms of simpler coefficients embedded in the determinant of a three by three matrix,
\begin{equation}
\begin{aligned}       
 J_{\mathcal{T}_{\kappa m_S}^\lambda}
      &=\frac{1}{b_1}
	\begin{vmatrix}
		&b_1&e_1&a_1&\\
		&b_2&e_2&a_2&\\
		&b_3&e_3&a_3&
	\end{vmatrix}.
\end{aligned}
\end{equation} The  coefficients appearing in this expression are given in the appendix. $J_{RG}$ transforms the input parameters from the GUT scale to the electroweak scale, and factorises as shown due to the supersymmetric non-renormalization theorem. The subscript $\lambda$ indicates that this parameter is kept constant in the derivatives.

As happens in going from Eq.~\ref{Eq:ptoO} to Eq.~\ref{bayestest} we can choose to work with the logarithms of parameters (as is natural if we choose logarithmic priors) so that we obtain a new factor in the denominator,
which is the inverse of the Jacobian with logarithms inserted inside the derivatives.  This gives us 
\begin{equation}
\begin{aligned}
\Delta_J^\textrm{CNMSSM}&=\left|\fp{\ln(m_Z^2,\tan\beta,\lambda)}{\ln(\kappa_0,m_{S_0}^2,\lambda_0)}\right| \\
 &= \frac{\kappa_0 m_{S_0}^2 \lambda_0}{m_Z^2 \tan\beta \lambda} J_{\mathcal{T}_0}^{-1}
\end{aligned}
\end{equation}

It is well known that the top quark Yukawa coupling can play a significant role in fine-tuning so we also considered this by extending the transformation to include the top quark mass and (unified) Yukawa coupling,  $(\kappa_0,m_{S_0}^2,\lambda_0,y_0) \rightarrow (m_Z^2,\tan\beta,\lambda,m_t)$.  Nonetheless as was already observed in the MSSM case \cite{Cabrera:2008tj,Cabrera:2009dm}, we found that all the derivatives, other than $\fp{m_t}{y_t}$, that involve $m_t$ and $y_t$ cancel, so this only changes the Jacobian by a single multiplicative factor of $\fp{m_t}{y_t}$.  Finally when logarithmic priors are chosen this factor will disappear entirely because  $\fp{\ln m_t}{\ln y_t} = 1$, and the Yukawa RGE factor $\fp{\ln y_t}{\ln y_0}$ is the same order one constant (at 1-loop) as in the CMSSM case so we neglect it. 
 
Therefore we write our NMSSM Jacobian based tuning measure as,   
\begin{equation}
	\Delta_J^\textrm{CNMSSM}=\left|\fp{\ln(m_Z^2,\tan\beta,\lambda,m_t^2)}{\ln(\kappa_0,m_{S_0}^2,\lambda_0,y_0^2)}\right|, \label{del_JCNMSSM}
\end{equation}
 with the additional transformation between $m_t$ and $y_0$ included to emphasise that we have also considered these, since the cancellation will prove to be rather important (in both the MSSM and NMSSM) when we compare against the Barbieri-Giudice tuning measure in the focus point (FP) region.  There we will show that due to this cancellation we do not see a large tuning penalty in the much discussed FP region\cite{Feng:2011aa,Feng:2012jfa,Feng:1999zg,Feng:1999mn}, which appears in the Barbieri-Giudice measure when one includes $y_t$ as a parameter.

The expression given here is formally the Jacobian which should be
used in the Bayesian analysis of any NMSSM model when $(\lambda_0,\kappa_0,m_{S_0}^2,y_0^2)$
are traded for $(m_Z^2,\tan\beta,\lambda,m_t^2)$.  At the same time
$\Delta_J^\textrm{CNMSSM}$ can be interpreted as a measure of the
naturalness of the NMSSM, which may be applied to the CNMSSM, the
general NMSSM and $\lambda$-SUSY scenarios.

Interestingly, as it was argued in the recent literature \cite{Ghilencea1311.6144}, the above Jacobians can also be considered to measure fine-tuning from a purely frequentist perspective.  In this context the same Jacobians appear as part of the likelihood function after one includes observables in $\chi^2$ which are related to the scale of electroweak symmetry breaking, such as the mass of the $Z$ boson.  Just as above, the variable transformation from these observables to fundamental parameters induces the Jacobian, which can be interpreted as a part of the likelihood that measures the sensitivity of the predicted electroweak scale to the fundamental parameters of the model.  Steep derivatives of the relevant observables with respect to the chosen fundamental parameters signal a strongly peaked likelihood function, indicating that $\chi^2$ drops off rapidly from the best fit value as those parameters are changed, which is of course indicative of high fine-tuning. The Bayesian perspective offers additional insight into the reasons we might dislike such behaviour in our likelihood functions, since in the frequentist case the actual best-fit $\chi^2$ does not suffer a penalty for any tuning observed in its vicinity, while in the Bayesian case there is a clear and direct penalty originating from the small prior--likelihood overlap that such behaviour implies.

 \section{Numerical analysis}
For our numerical analysis we use \texttt{SOFTSUSY 3.3.5} for the MSSM
\cite{Allanach:2001kg}, and \texttt{NMSPEC} \cite{Ellwanger:2006rn} in
\texttt{NMSSMTools 4.1.2} for the NMSSM. \texttt{Next-to-Minimal
  SOFTSUSY} \cite{Allanach:NMSOFTSUSY} was still in development during
this analysis but was used to cross check the spectrum for certain
points.  \texttt{MultiNest 3.3} was used for scanning
\cite{Feroz:2008xx,Feroz:2007kg}. Both spectrum generators used here
provide $\Delta_{BG}$ with renormalization group flow improvement.
For $\Delta_{BG}$ in the CMSSM we include individual sensitivities,
$\Delta_{BG}(p_i)$, for the set of parameters $M_0, M_{1/2}, A_0, \mu,
B, y_t$. For the CNMSSM we use the set $M_0, M_{1/2}, A_0, \lambda,
\kappa, y_t$.

First we examine how the tuning measures vary with $M_0$ and
$M_{1/2}$, without requiring a $125$ GeV Higgs.  We fix $\tan
\beta=10$, where the extra NMSSM F-term contribution is small, but
there is interesting focus point (FP) behavior
\cite{Feng:2011aa,Feng:2012jfa,Feng:1999zg,Feng:1999mn}.  Previous
studies \cite{Kowalska:2012gs} show that large and negative $A_0$ is
favoured, so to simplify the analysis here and throughout we
choose\footnote{We checked that with alternative $A_0$ choices the
  behaviour is similar. The main difference is with the Higgs masses
  where a large and negative $A_0$ was chosen to increase the lightest
  Higgs mass.} $A_0=-2.5$ TeV.

\begin{figure}[h!]
        \includegraphics[width=0.5\textwidth]{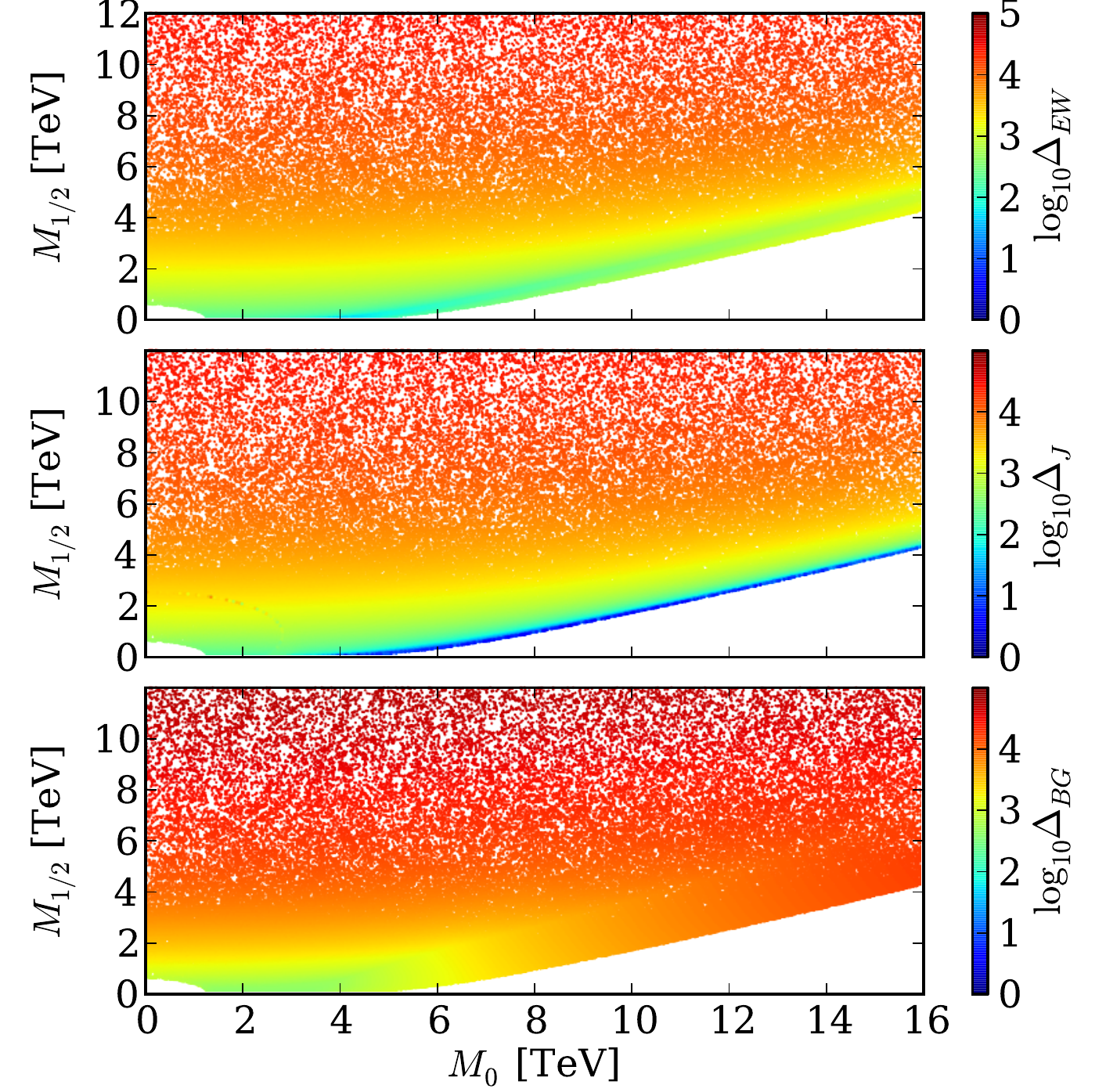}
	\caption{Fine tuning measures $\Delta_{BG}$ (top), $\Delta_J$
          (middle), $\Delta_{EW}$ (bottom) in the $M_0$ vs. $M_{1/2}$
          plane for $A_0=-2.5$ TeV, $\tan\beta=10$ and sgn($\mu$)=1 in
          the CMSSM.  The color code quantifies the value of
          $\Delta_{EW}$ and $\Delta_J$.  Since $\Delta_{BG}$ is
          dominated by the $\mu$ derivative it is low in the small
          $M_0$ and $M_{1/2}$ region.  Although $\Delta_{BG}$, by
          definition, is formally part of $\Delta_J$ the numerical
          behaviour of the latter is similar to that of $\Delta_{EW}$.
          All massive parameters are in GeV unit.  No experimental
          constraints applied except that the lightest supersymmetric
          particle is electrically neutral and the EWSB condition is
          satisfied.
	\label{delbg_j}}
\end{figure}

 The results for the CMSSM are shown in FIG.~\ref{delbg_j}.  The value
 of $\Delta_{EW}$ is governed by the $m_{H_u}^2$ and $\mu^2$
 contributions since $m_Z^2/2 \approx -\overline{m}_{H_u}^2-\mu^2$,
 where $\overline{m}_{H_u}^2$ includes the radiative corrections.  In
 general $\Delta_{EW}$ is dominated by $\mu^2$, while the crossover to
 the $m_{H_u}^2$ dominance occurs in the vicinity of the EWSB
 boundary.

For this measure there is low fine-tuning even at large $M_0$. This
may seem counterintuitive, but for $\tan\beta = 10$ at large $M_0$ we
are close to a FP region.  In this region the dependence on $M_0$
which appears from RG evolution of $m_{H_u}$ vanishes.  For example in
the CMSSM semi-analytical solution to the renormalisation group
equations (RGEs), \begin{eqnarray} m_{H_u}^2&=&c_1 M_0^2 + c_2
  M_{1/2}^2 + c_3 A_0^2 + c_4 M_{1/2} A_0, \label{Eq:mhusq}
\end{eqnarray}
the coefficients $c_i$ are functions of Yukawa and gauge couplings,
and $\tan \beta$ and $c_1$ can be close to zero.  Such regions then
appear to have low fine-tuning even with large $M_0$ since the small
size of $c_1$ means there is no need to cancel the large $M_0$ in
Eq.~(\ref{Eq:mzsq}) to obtain the correct $m_Z^2$.

In $\Delta_{BG}$, however, the sensitivity to the top quark Yukawa
coupling is included.  Since the RG coefficients depend on this Yukawa
coupling, the large stop corrections from the RGEs that feed into
$m_{H_u}^2$ lead to a large $\Delta_{BG}(y_t)$ even in the focus point
region.  $\Delta_{EW}$ is not sensitive to this effect since it does
not take into account such RG effects.

\begin{figure}[h!]
        \includegraphics[width=0.5\textwidth]{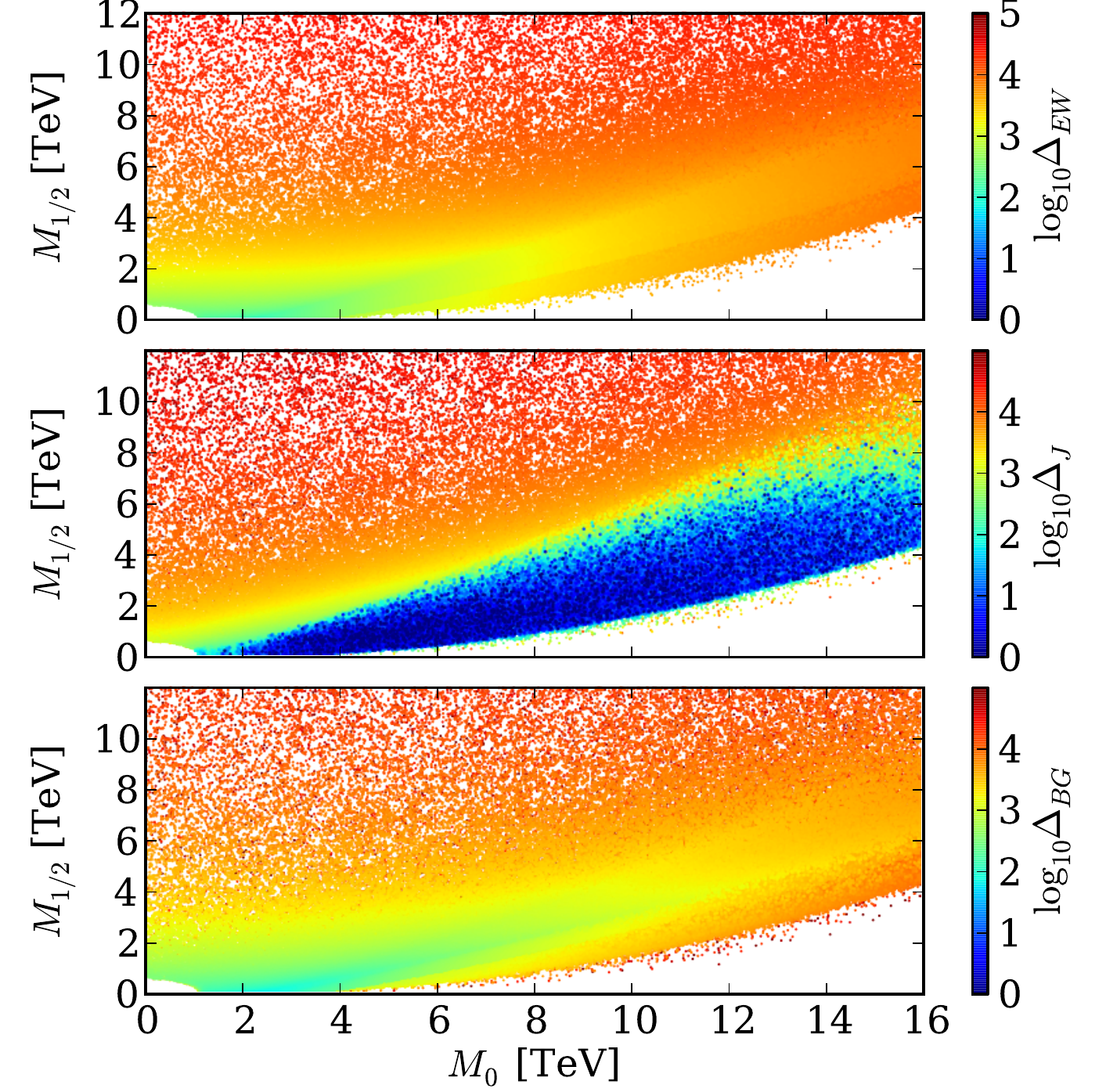}
	\caption{Same as FIG.~\ref{delbg_j} except for the constrained
          NMSSM. $A_{0,\kappa,\lambda}=-2.5$~TeV and $\tan\beta=10$
          are assumed. $\lambda$ is sampled from the range [0,0.8].
	\label{bg_vs_j_nm}}
\end{figure}

Interestingly $\Delta_J^\textrm{CMSSM}$ exhibits similar behavior to
$\Delta_{EW}$ despite containing derivatives from $\Delta_{BG}$.  This
is because $\Delta_J^\textrm{CMSSM}$ does not contain the derivative
of $m_Z$ with respect $y_t$.  When one computes the Jacobian for
Eq.~(\ref{del_JCMSSM}) the derivative of $y_t$ with respect to $m_Z$
cancels out, leaving only the derivatives
$\fp{\mu}{M_z}\fp{B\mu}{t}\fp{y_t}{m_t}$ in the Jacobian.  As a result
$\Delta_J$ in the MSSM can remain small in the focus point region.
 
Fine tuning measures for the CNMSSM are shown in
FIG.~\ref{bg_vs_j_nm}.  Here $\Delta_J^\textrm{CNMSSM}$ is defined by
Eq.~(\ref{del_JCNMSSM}) and $\Delta_{BG}$ is defined by
Eq.~(\ref{del_BG}), while $\Delta_{EW}$ is defined the same as for
the MSSM.  The parameter $\mu$ dominates electroweak tuning, $\Delta_{EW}$,
 throughout the $M_0$ vs. $M_{1/2}$ plane. Since $\mu$ values and
 related derivatives are similar in the CMSSM and CNMSSM the fine-tuning 
 measures are qualitatively similar for the two models.

As in the CMSSM the Jacobian derived tuning $\Delta_J$ increases with
$M_{1/2}$, as anticipated since for large $M_{1/2}$ large cancellation
is required to keep $m_Z$ light. Again though at large $M_0$
$\Delta_J$ can still be low seeming to favour this FP region, which is
a result of the same cancellation as happened in the MSSM case
occurring in our new NMSSM Jacobian.

 Interestingly the region where the tuning can be very low extends
 further in the NMSSM.  Note this is not a result of raising the Higgs
 mass with $\lambda$ since we impose no Higgs constraint yet and have
 large $\tan\beta$.  However $\lambda$ is varied across the plane and
 affects the EWSB condition and the renormalization group evolution.
 However since the number of parameters are different in the CNMSSM
 and CMSSM, to determine whether the CNMSSM is preferred over the
 CMSSM, we have to compare Bayesian evidences.

\begin{figure}
	\includegraphics[width=0.45\textwidth]{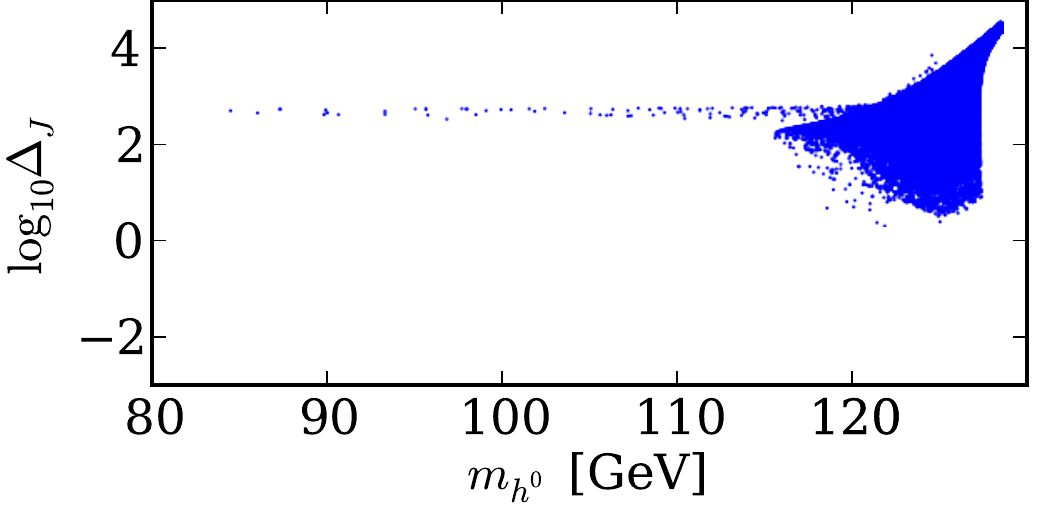}
	\includegraphics[width=0.5\textwidth]{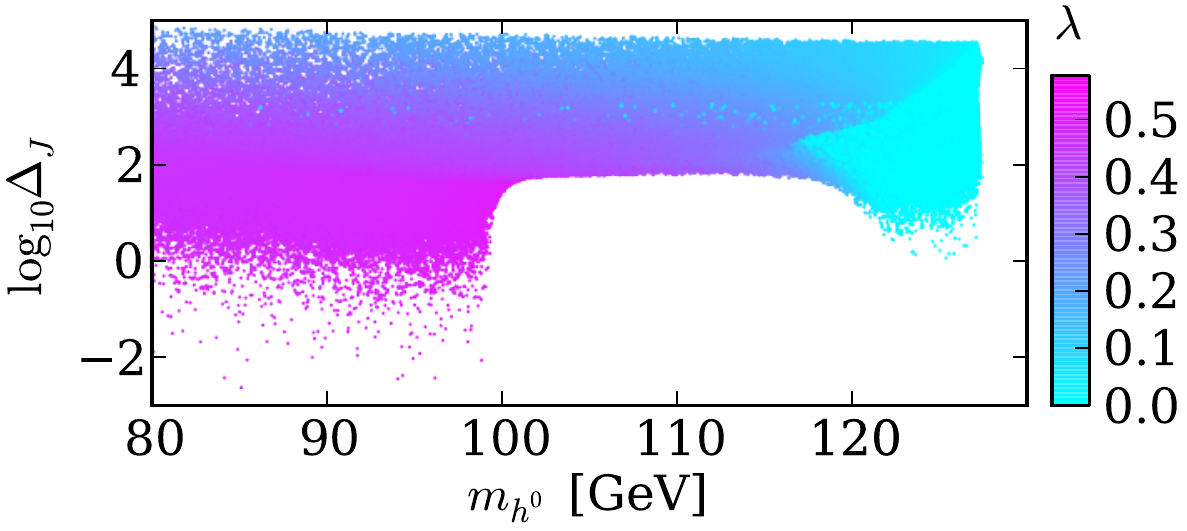}
	\caption{Fine tuning with respect to $m_{h^0}$ for the CMSSM (upper) and CNMSSM (lower). $A_0=-2.5$~TeV and $\tan\beta=10$ for both models.}
	\label{delj_mh}
\end{figure}

Since the focus point region allows small $\Delta_{EW}$ and
$\Delta_{J}$ in the large $M_0$ region it is possible to have a
relatively heavy lightest Higgs and small $\Delta_{J}$.  This is
illustrated in FIG.~\ref{delj_mh}.  Note also that in the NMSSM case
there is no tuning preference for large $\lambda$ since the new F-term
contribution goes like $\lambda^2 v^2 \sin^22\beta$ and is therefore
suppressed at large $\tan \beta$.  Nonetheless in the focus region in
both the CMSSM and CNMSSM one can have a $125$ GeV without an enormous
penalty from effective prior weighting $\Delta_J$. 

However the lowest tuning is when $M_{1/2}$ is
smallest and this region is strongly constrained by squark and gluino
searches.  The important message, nonetheless, is that the Higgs mass
measurement has a low impact on naturalness in the focus point region.
Therefore the effect of the Higgs mass measurement may not be as
severe on our degree of belief as we would expect from $\Delta_{BG}$,
even in the MSSM.
A caveat to this optimistic statement is that from looking at
$\Delta_J$ alone one cannot know if the focus point scenarios will be
suppressed by other factors in the full Bayesian analysis. This can
only be determined by carrying out that analysis.

\begin{figure}
	\includegraphics[width=0.45\textwidth]{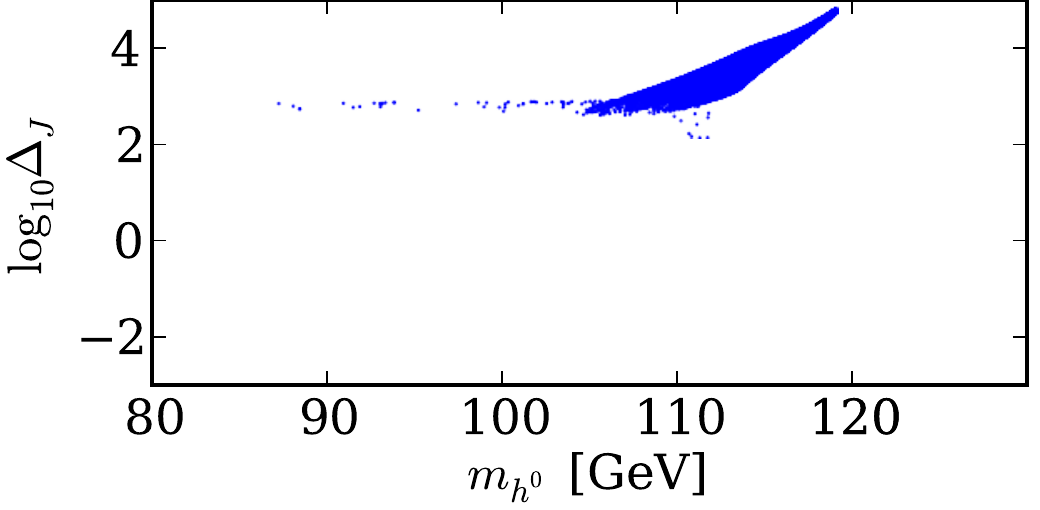}
	\includegraphics[width=0.5\textwidth]{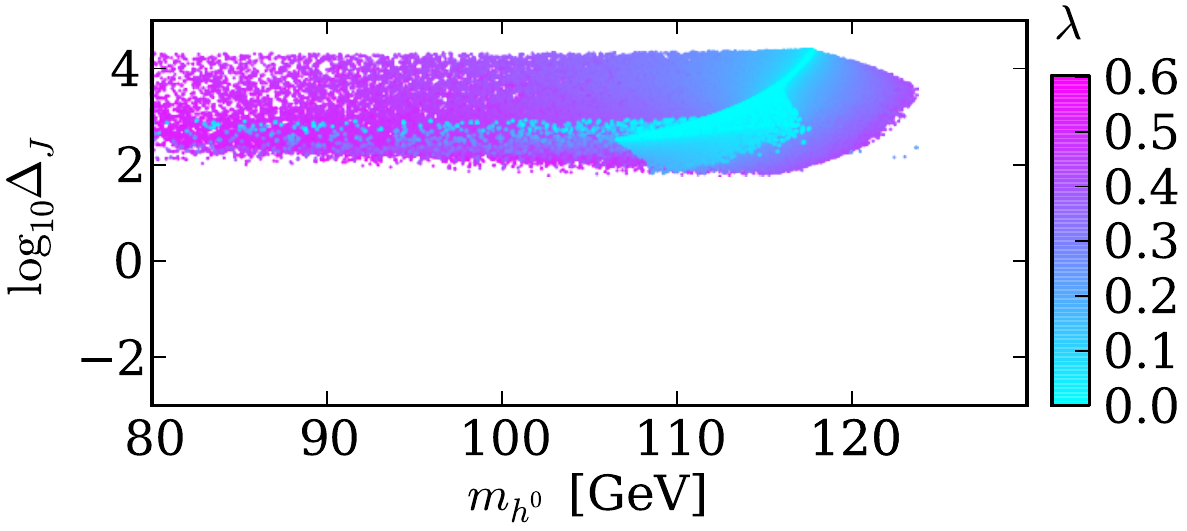}
	\caption{Fine tuning with respect to $m_{h^0}$ for the CMSSM (upper) and CNMSSM (lower). $A_0=-2.5$~TeV and $\tan\beta=3$ for both models.}
	\label{DelJ-lowtb}
\end{figure}

Away from this special FP region the Higgs mass measurement has a
large impact and the extra $F$-term of the NMSSM can play a vital
role.  In Fig.~\ref{DelJ-lowtb} we compare the Higgs mass against 
fine-tuning for $\tan\beta =3$ in both the CMSSM and the CNMSSM.  Here the
extra NMSSM F-term can give a larger contribution to the SM-like Higgs
mass and it is precisely this effect which leads to expectations of
increased naturalness in the NMSSM.

In the MSSM the tree level upper bound reduces rapidly at small $\tan
\beta$. Therefore we do not find any CMSSM solutions with a lightest
Higgs mass above $120$ GeV in FIG.~\ref{DelJ-lowtb}. The maximum
achievable mass of the lightest Higgs has $\Delta_J \approx 10^5$.
By comparison the same mass for the lightest Higgs in the CNMSSM can
be achieved with $\Delta_J$ between $10^2-10^3$. So according to
the naturalness prior measure $\Delta_J$ the tuning is reduced
compared to the CMSSM for heavier Higgs masses.

 Nonetheless
for $m_{h^0} > m_Z$ on contours of fixed $\lambda$,
 $\Delta_J$
increases with the lightest Higgs mass and the minimum
 $\Delta_J$
starts increasing significantly when the lightest Higgs
 mass is
pushed above $115$ GeV.  As expected the largest Higgs masses
 are
found for sizable $\lambda$.  This demonstrates that for the new
Jacobian naturalness measure for the NMSSM the additional F-term
contribution in the NMSSM really does decreasing fine-tuning of the
model as one increases $\lambda$, strongly supporting previous that this
mechanism can reduce fine-tuning in the low $\tan \beta$ region of
the NMSSM.
 
 However for the $\tan\beta=3$ slice it is still hard
to achieve a
 $125$ GeV lightest Higgs mass in such strongly
constrained
 scenarios. $\lambda$ does not reach the perturbative
limit, with
 $\lambda \leq 0.6$.  Unlike the MSSM, $A$ terms play an
important
 role in the EWSB condition. Since $B=A_\lambda+\kappa s$,
$A_\lambda$ restricts the parameter space by the tachyonic CP-odd
mass constraint.  Further, $A_\kappa$ also affects the EWSB
condition through the validity of the global minimum\footnote{For
  example in the large $s$ limit this requires
  $A_\kappa^2>8m_S^2$. This must be satisfied simultaneously with,
  $A_\kappa=A_0$ and the minimisation condition involving $m_S^2$}.
While a $125$ GeV Higgs in constrained versions is difficult to
achieve, it is easier in the unconstrained
NMSSM\cite{King:2012tr,King:2012is} . Therefore a detailed analysis
of
 the multidimensional unconstrained NMSSM is required, and this
will
 be presented in our companion paper \cite{Monash_Adelaide}
where we
 consider both the perturbative NMSSM and $\lambda$-SUSY
scenarios.
 
\begin{figure}
	\includegraphics[width=0.5\textwidth]{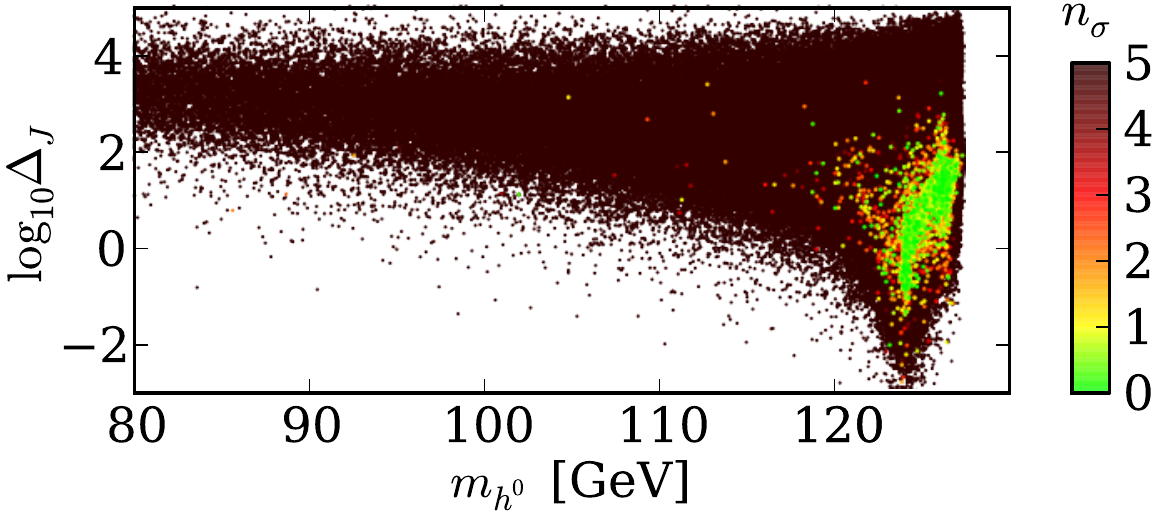}
	\includegraphics[width=0.5\textwidth]{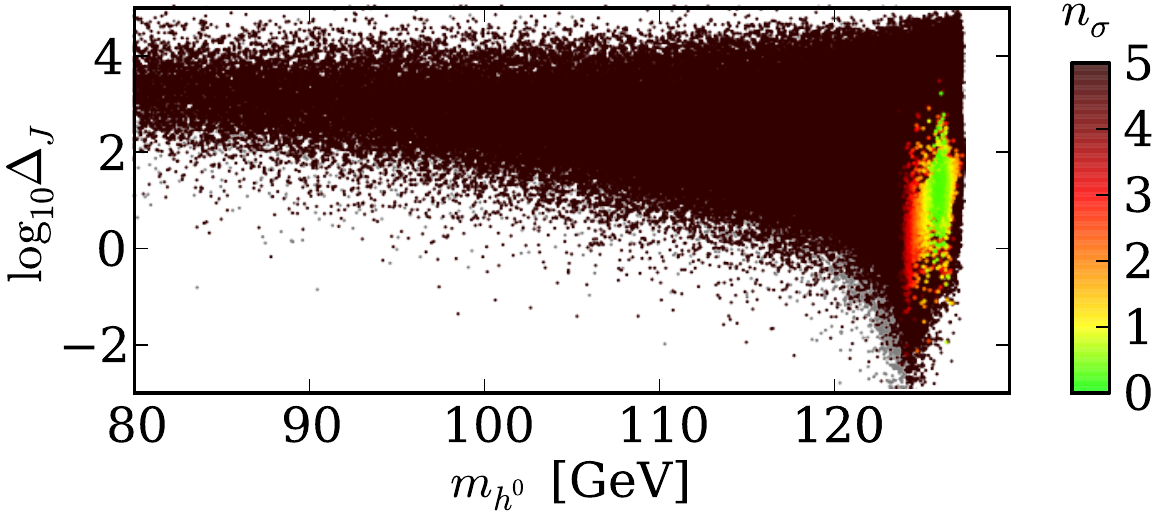}
	\caption{Fits to various observables in the framework of the slightly relaxed CNMSSM for fixed values of $A_0=-2.5$~TeV and $\tan\beta=10$.  $A_\lambda$ and $A_\kappa$ are allowed to vary independently from $A_0$.  TABLE \ref{tab:observables} shows the experimental values of the observables that were used in this fit.  The top frame is the fit to the relic density alone while the bottom is the two observable combined fit.  The statistical significance with which each model point can be rejected is given in units of $\sigma s$.  As the figure shows on the $A_0=-2.5$~TeV and $\tan\beta=10$ hypersurface a good fit to both observables can be obtained for the low Jacobian tuning of $\Delta_J\sim1$.}
	\label{fig:fits}
\end{figure}

FIG.~\ref{fig:fits} shows fits to various observables in the framework of the slightly relaxed CNMSSM for fixed values of $A_0=-2.5$~TeV and $\tan\beta=10$.  For these scans $A_\lambda$ and $A_\kappa$ are allowed to vary independently from $A_0$.  We decouple $A_\lambda$ and $A_\kappa$ from $A_0$ to easily obtain a neutralino relic density and a lightest Higgs mass which simultaneously satisfy the experimental constraints.  TABLE \ref{tab:observables} shows the experimental values of the observables that were used in the fit shown in FIG.~\ref{fig:fits}.  The neutralino relic density is required to match the dark matter relic density as measured by Planck \cite{Ade:2013ktc}.  For the lightest Higgs mass we use the PDG combined value \cite{PDG:2013}. PDG combined limits are used to constain the sparticle masses, except for the squark and gluino masses; in this case we take the strongest currently listed PDG limits, even though these do not directly apply to the model under consideration, in order to be conservative. The constraints on rare B decays are taken from LHCb\cite{Aaij:2013aka} and HFAG\cite{Amhis:2012bh}. All constraints are implemented as Gaussian likelihoods except where a limit is indicated, in which case a hard cut is applied.

\begin{table}
\caption{Experimental values of the observables that were used in the fit shown in FIG.~\ref{fig:fits}. \label{tab:observables}}
\begin{ruledtabular}
\begin{tabular}{lr}
Observable        &                  Experimental value  \\
\hline
$\Omega_{DM} h^2$                                    & $0.1187\pm0.0017$ \cite{Ade:2013ktc} \\
$m_h$                                                & $125.9\pm0.4$ GeV \cite{PDG:2013} \\
$\mathrm{BR}\left( B_s \rightarrow \mu^+\mu^-\right)$ & $(2.9\pm1.1) \times 10^{-9}$ \cite{Aaij:2013aka}
\\
$\mathrm{BR}\left( b \rightarrow s \gamma \right)$    & $(343\pm21\pm7) \times 10^{-6}$ \cite{Amhis:2012bh}
\\ 
$\mathrm{BR}\left( B \rightarrow \tau\nu \right)$     & $(114\pm22) \times 10^{-6}$ \cite{Amhis:2012bh}
\\
$m_{\tilde{\chi}^0_1}$                               & $> 46$ GeV \cite{PDG:2013}\\
$m_{\tilde{\chi}^\pm_1}$                             & $> 94$ GeV if $m_{\tilde{\chi}^\pm_1} - m_{\tilde{\chi}^0_1} > 3$ GeV\cite{PDG:2013}\\
$m_{\tilde{q}}$                                      & $> 1.43$ TeV \cite{PDG:2013}\\
$m_{\tilde{g}}$                                      & $> 1.36$ TeV \cite{PDG:2013}\\
\end{tabular}
\end{ruledtabular}
\end{table}


The top frame of FIG.~\ref{fig:fits} is the fit to the relic density alone while the bottom is the two observable combined fit.  The statistical significance with which each model point can be rejected is given in units of $\sigma s$.  These significances correspond to local p-values, computed assuming the observables' best fit values are normally distributed with the specified standard deviation.  To be conservative, no additional theoretical uncertainty is included in the fit.  As the figure shows on the $A_0=-2.5$~TeV and $\tan\beta=10$ hypersurface a good fit to both observables can be obtained for the low Jacobian tuning of $\Delta_J\sim1$. 

\section{Conclusions}

In this work we presented Bayesian naturalness priors to quantify fine-tuning in the (N)MSSM.  These priors emerge automatically during model comparison within the Bayesian evidence.

We compared the Bayesian measure of fine-tuning ($\Delta_J$) to the Barbieri-Giudice ($\Delta_{BG}$) and ratio ($\Delta_{EW}$) measures.  Even though the Bayesian prior is closely related to the Barbieri-Giudice measure, the numerical value of the Bayesian measure reproduces important features of $\Delta_{EW}$. Both $\Delta_{EW}$ and $\Delta_J$ are low in FP scenarios.

Our numerical analysis is limited to fixed ($A_0,\tan\beta$) slices of the constrained parameter space.  For these slices we show that, according to the naturalness prior, the constrained version of the NMSSM is less tuned than the CMSSM.  This statement, however, has to be confirmed by comparing Bayesian evidences of the models.  The complete parameter space scan and the full Bayesian analysis for the NMSSM is deferred to a later work \cite{Monash_Adelaide}.

\section{Acknowledgements}

This research was funded in part by the ARC Centre of Excellence for Particle Physics at the Tera-scale, and in part by the Project of Knowledge Innovation Program (PKIP) of Chinese Academy of Sciences Grant No. KJCX2.YW.W10.  The use of Monash Sun Grid (MSG) and the Multi-modal Australian ScienceS Imaging and Visualisation Environment (www.MASSIVE.org.au) is also gratefully acknowledged.  DK is grateful to Xerxes Tata for the useful discussion.  PA thanks Roman Nevzorov and A.~G.~Williams for helpful comments and discussions during the preparation of this manuscript.

\begin{appendix}
\appendix
\section{Appendix: Jacobian entries}
The entries appearing in the Jacobian $ J_{\mathcal{T}_{\kappa m_S}}$ in Eq.~\ref{Eq:Jac_kapms} are given in this appendix.
\ba a_1 &=& - \kappa A_\kappa - 4 \kappa^2 s - \frac{\lambda A_\lambda v^2 s_{2\beta}}{2 s^2} \hspace{4mm} \\
  a_{2} &=&  \lambda \kappa v^2 \fp{ s_{2\beta}}{t_\beta} + \frac{\lambda A_\lambda v^2 }{2 s } \fp{ s_{2\beta}}{t_\beta} \hspace{4mm} \\
a_{3}& =& - \frac{\lambda^2 v^2}{M_Z^2} + \lambda \kappa s_{2\beta}\frac{ v^2}{M_Z^2} + \frac{\lambda A_\lambda v^2 s_{2\beta}}{2 s M_Z^2}
 \ea
\ba b_1& =& - \frac{\lambda}{s} \hspace{1cm}    b_2 = \frac{1}{2\lambda s^2}\frac{2t_\beta}{(t_\beta^2-1)^2} (m_{H_u}^2-m_{H_d}^2) \hspace{1cm}\\
  b_3 &=&  -\frac{1}{4\lambda s^2} \ea
\ba
e_1 & =& -\frac{2\lambda s\sin{2\beta}-(A_\lambda+2\kappa s)}{s^2} \hspace{1cm}\\
e_2  &=& - \frac{1-t_\beta^2}{t_\beta(1+t_\beta^2)}\frac{A_\lambda+\kappa s}{s} \hspace{3.7cm} \\
e_3 &=& -\frac{\lambda\sin{2\beta}}{\bar{g}^2s^2}  \ea

\end{appendix}

\end{document}